# Next-Generation Geodesy at the Lunar South Pole: An Opportunity Enabled by the Artemis III Crew.

Vishnu Viswanathan[1,2], Erwan Mazarico[1], Stephen Merkowitz[1], Xiaoli Sun[1], T. Marshall Eubanks[3], David E. Smith[4]
[1] NASA Goddard Space Flight Center, Greenbelt, MD
[2] University of Maryland, Baltimore County, MD
[3] Space Initiatives Inc. Palm Bay, FL
[4] Massachusetts Institute of Technology, Cambridge, MA

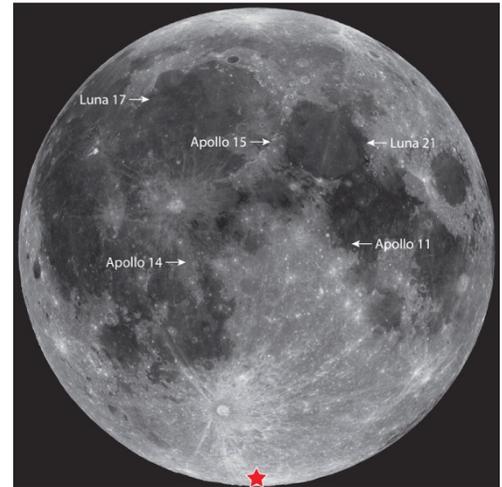

Fig.1: An opportunity to expand LLR science experiment to the lunar south pole through the Artemis III mission.

Lunar retro-reflector arrays (LRAs) consisting of corner-cube reflectors (CCRs) placed on the nearside of the Moon during the Apollo era have demonstrated their longevity, cost-effectiveness, ease of deployment, and most importantly their interdisciplinary scientific impact through the ongoing lunar laser ranging (LLR) experiment. The human exploration of the lunar south polar region provides a unique opportunity to build on this legacy and contribute to the scientific return of the Artemis, for many decades to come. Here we outline the extended science objectives realizable with the deployment of geodetic tracking devices by the Artemis III crew.

*Dynamical signatures of a lunar solid inner core.* Geophysical parameters estimated from Apollo seismic data analyses[1,2] and GRAIL gravity data analyses[3,4], result in a range of lunar interior models[5] that permit a Moon with or without a solid inner core. Gravitational signatures (time variations of the spherical harmonic coefficients $C_{21}$ and $S_{21}$) require i) a non-zero angle between the inner core and the mantle equators to the ecliptic; ii) a non-spherical boundary at the interface of the inner and fluid outer core, both of which are plausible conditions[6,7]. However, the GRAIL data analyses did not report such weak gravitational signatures. LLR data analyses offer high sensitivity to subtle variations in the Moon's rotational dynamics potentially extending up to the depths of an expected solid inner core. A recent study[8] reports an inner core size > 200 km in radius may contribute to as much as a few thousandths of the LLR observed mantle precession angle of 1.543°. Lunar orientation derived from fits to LLR data are presently known to a few tens of milliarcseconds, offering an independent method to detect the solid inner core. Also, non-sphericity within the lunar interior layers is sensitive to LLR. The polar shape of the core-mantle boundary is detected using present LLR data[9] but the equatorial shape ($\beta$) remains undetected due to limited geometry and range accuracy. An estimate of $\beta$ will help understand lunar dynamo mechanisms in the past driven by impacts and tidal instabilities[10]; and, contribute to our knowledge of the hydrostatic state of the lunar core. A non-zero $\beta$ will also modify the free core nutation and induce a new proper mode.

*Constraints from lunar tidal deformation* can be parameterized using vertical ($h_2$) and horizontal ($l_2$) tidal-displacement Love numbers. The estimates from the analysis of LLR data[5] and altimetric crossover data[11] give a value of $h_2$ of 0.048(6) and 0.037(3) respectively. The LLR values are only slightly larger, however, the differences in these estimates may be significant enough to differentiate between end-member lunar interior models[5]. The ability to separate end-member models based on the observed tidal responses would help understand the true lunar interior structure. The small size of the lunar core and an uncertain density profile makes this a challenge and thus necessitate more precise measurements. The recovery of lunar $h_2$ using LLR can also be improved with a better spatial network of retroreflectors. The largest monthly terms have radial displacements of 11.2 cm and 4.5 cm at the lunar equator and poles respectively[12]. LLR retroreflectors placed at the lunar south pole will help decorrelate the retroreflectors' body-fixed coordinates and lunar orientation parameters recovered using the current network. A radio beacon[13] placed at the lunar south pole would help Earth-based VLBI stations to observe the reduced polar tidal displacements as precise angular measurements, complementing the LLR technique. Alternatively, a single small LRA placed near the lunar rotational axis would serve as an important surface marker seen every orbit by a typical polar-orbiting lidar-enabled spacecraft (e.g., LRO-LOLA). Small LRAs (5 cm base diameter, 20g in mass)[14] built under NASA's Commercial Lunar Payload Service (CLPS) program are to be carried on all CLPS landers and would help address this, also serving as fiducial markers on the lunar surface to support navigation and geolocation, and potentially establishing improved and independent direct ties between lunar reference frames (LLR-derived principal axis frame with LRO products based in the mean-Earth frame). The current sensitivity provided by LLR for the independent estimation of the lunar horizontal tidal-displacement love number ($l_2$) is insufficient due to the present geometry allowed by current LLR LRA network. Thus, lunar $l_2$ determinations from LLR analysis remain model dependent. A south polar Earth-facing, Earth-visible, LLR retroreflector deployed by the Artemis crew will provide the required complement to the current LLR network, with its large latitudinal separation from the existing network of LRAs. Site selection for surface deployment must consider average Earth-visibility (mostly < 50% within 6° of the lunar south pole[15]) resulting from lunar south polar topography[16]. A dominant random uncertainty per returned photon stems from the tilt of the large LRA that oscillates with Moon's libration (since the LRAs have a fixed tilt tied to the crust). This creates an ambiguity in identifying the individual CCR that contributed to the return signal, resulting in a greater pulse spreading. Multiple smaller CCRs separated by few tens of cm illuminated together by a typical few km LLR spot size will be advantageous to overcome present LRA range data precision limitation from pulse spreading[17] and would potentially enable a differenced observable sensitive to relative local deformations.

*Precision tests of fundamental physics.* The five decades of LLR data suffer from a non-uniform distribution in lunar phase, due to the impact of thermal gradients on the CCR's optical properties. This degrades the precision on the tests of fundamental physics using LLR, such as the equivalence principle, because the absolute maximum of the violation signal occurs at times when the LLR data are scarce from a heated CCR[18,19]. Furthermore, the weak signals from LLR limit present operations to a select few stations on Earth due to the single-photon detection regime imposed by factors such as telescope aperture, laser power, beam divergence and throughput-loss from two-way transmission through the Earth's atmosphere, detection efficiency, etc. An active laser transponder[20] would have a significant gain in signal strength, high accuracy and complementary geometry that will expand the LLR detection capability to the wider SLR network (45+ stations distributed globally) compared to just a few stations in a small northern-latitude band. Asynchronous laser transponders with accurate clock referencing have demonstrated capabilities[21]. Their application could also extend the lessons learned from LLR-derived science to interplanetary distances[22] and they can support high-precision requirements[23] for future planetary missions[24,25] and time transfer applications[26,27].

We recommend the Science Definition Team to consider the Artemis crew deployment of a combination of active and passive geodetic devices, like active laser transponders, radio beacons (to support differential transverse measurements) and/or several single CCRs for 2-way ranging from Earth (LLR). These will contribute to the following science objectives:
- to enable high-precision long-term monitoring of the lunar orbit, orientation, rotation and tidal properties to understand dynamical signatures from the deep lunar interior;
- to enable the long-term maintenance of lunar ephemeris for future high-precision navigation and to complement solutions of Earth Orientation Parameters;
- to enable stringent tests of fundamental physics using the Earth and the Moon as test bodies;
- to establish independent high-accuracy ties of the lunar body-fixed frame to the ICRF using differential VLBI with radio beacons.